\documentclass[12pt]{iopart}
\usepackage{graphicx}
\usepackage{float}
\usepackage{color}
\usepackage{nameref}
\usepackage[normalem]{ulem}
\usepackage{amssymb}
\usepackage[sort&compress,numbers]{natbib}
\usepackage{hyperref}
\hypersetup{
	colorlinks,
	linkcolor={blue},
	citecolor={blue},
	urlcolor={blue}
}

\begin{document}


\newcommand{\change}[1]{\textcolor{black}{#1}}
\newcommand{\remove}[1]{\textcolor{black}{}}

\title[Strategy for accurate thermal biasing at the nanoscale]{Strategy for accurate thermal biasing at the nanoscale}

\author{Artem O. Denisov\footnote{Present address:
Department of Physics, Princeton University, Princeton, New Jersey 08544, USA}}
\address{Institute of Solid State Physics, Russian Academy of Sciences, 142432 Chernogolovka, Russian Federation}
\address
{Moscow Institute of Physics and Technology, Dolgoprudny, 141700 Russian Federation}
\author{Evgeny S. Tikhonov}
\address
{Institute of Solid State Physics, Russian Academy of Sciences, 142432 Chernogolovka, Russian Federation}
\address
{National Research University Higher School of Economics, 20 Myasnitskaya Street, Moscow 101000, Russian Federation}
\author{Stanislau U. Piatrusha}
\address
{Institute of Solid State Physics, Russian Academy of Sciences, 142432 Chernogolovka, Russian Federation}
\author{\change{Ivan N. Khrapach}}
\address{Institute of Solid State Physics, Russian Academy of Sciences, 142432 Chernogolovka, Russian Federation}
\address{Moscow Institute of Physics and Technology, Dolgoprudny, 141700 Russian Federation}
\address{Russian Quantum Center, 121205 Skolkovo, Moscow, Russian Federation}
\author{Francesco Rossella, Mirko Rocci, Lucia Sorba}
\address
{NEST, Istituto Nanoscienze – CNR and Scuola Normale Superiore, Piazza S. Silvestro 12, I-56127 Pisa, Italy}
\author{Stefano Roddaro}
\address
{NEST, Istituto Nanoscienze – CNR and Scuola Normale Superiore, Piazza S. Silvestro 12, I-56127 Pisa, Italy}
\address
{Department of Physics "E.Fermi", Universita di Pisa, Largo Pontecorvo 3, I-56127 Pisa, Italy}
\author{Vadim S. Khrapai}
\address
{Institute of Solid State Physics, Russian Academy of Sciences, 142432 Chernogolovka, Russian Federation}
\address
{National Research University Higher School of Economics, 20 Myasnitskaya Street, Moscow 101000, Russian Federation}
	
\begin{abstract}
{\color{black} We analyze the benefits and shortcomings of a thermal control in nanoscale electronic conductors by means of the contact heating scheme. Ideally, this straightforward approach allows one to apply a known thermal bias across nanostructures directly through metallic leads, avoiding conventional substrate intermediation.} We show, by using the average noise thermometry and local noise sensing technique in InAs nanowire--based devices, that a nanoscale metallic constriction on a $\rm SiO_2$ substrate acts like a diffusive conductor with  negligible electron-phonon relaxation and  non-ideal leads. The non-universal impact of the leads on the achieved thermal bias --- which depends on their dimensions, shape and material composition --- {\color{black} is hard to minimize, but is possible to accurately calibrate in a properly designed nano-device. Our results allow to reduce the issue of the thermal bias calibration to the knowledge of the heater resistance and pave the way for accurate thermoelectric or similar measurements at the nanoscale.} \end{abstract}
\section*{Introduction}

 Managing nanoscale electronic devices out of thermal equilibrium is an outstanding problem both from the fundamental~\cite{Dubi2009,Ventra2016} and applied perspectives~\cite{Snyder2008}. Unlike cooling the electrons down by refrigeration~\cite{RevModPhys.78.217,Muhonen2012}, raising their temperature above the thermal bath is achieved easily by Joule heating of the whole device~\cite{Roukes1985}, of a part of it~\cite{PhysRevLett.81.2982,Tikhonov2016,Menges2016} or of the nearby substrate~\cite{Shi2003,Zuev2009,doi:10.1021/nl101505q,doi:10.1021/nl401501j,Roddaro2013,doi:10.1021/nl304619u,PhysRevLett.120.177703}. Accurate control of the generated thermal bias, which is an obvious prerequisite for a quantitative measurement in thermoelectric (TE) and other similar experiments~\cite{Seki2015,RouraBas2018}, remains a separate complex problem. \change{This is especially true for the electronic transport at nanoscale, when the dimensions of the device under test are small compared to the inelastic length scales, whereas the available current leads are too small to be treated as ideal macroscopic reservoirs.}

Various non-equilibrium control tools are capable of measuring the temperature of the electronic~\cite{RevModPhys.78.217,Wang2018,PhysRevLett.81.2982,Tikhonov2016} and lattice sub-systems~\cite{Yazji2015}, including spatially resolved~\cite{Doerk2010,Menges2016,Weng2018,WengScience2018}, time resolved~\cite{Schmidt2004,Gasparinetti2015} and energy resolved~\cite{PhysRevLett.79.3490,7985929} approaches. However, the accuracy of the control is strongly outweighed by the tools' complexity. Moreover, the modeling of the heat balance at nanoscale is often not reliable, for the {\it a-priori} unknown electron-phonon coupling~\cite{Wang2018}, thermal contact resistance~\cite{Shi2003,0957-4484-26-38-385401, PhysRevB.83.205416,Hochbaum2008,Yazji2016} and lattice thermal conductivity~\cite{doi:10.1063/1.1616981,Boukai2008,PhysRevB.83.205416}. For instance, as demonstrated recently by some of us, below $\sim40$\,K the electron and lattice systems are practically decoupled in InAs nanowires (NWs)~\cite{Tikhonov2016}. {\color{black} As a result, the substrate heating is inefficient for thermal biasing~\cite{0268-1242-31-10-104001} manifesting a failure of such an approach in this temperature range.} Not to mention that the thermal bias calibration by means of the metallic resistance thermometry~\cite{doi:10.1021/nl304194c} loses its sensitivity within the residual resistance range at low temperatures and is complicated for heaters shorter than the electron-phonon relaxation length.  These obstacles are easily overcome using a contact heating scheme accompanied by a primary calibration of the thermal bias via noise thermometry~\cite{PhysRevLett.81.2982}. In this case, a nearly ideal thermal contact between the electronic systems of a uniform current biased metallic heater and the nano-device is achieved by an ohmic contact of negligible resistance. Recently, this approach was successfully applied to TE measurements in  individual InAs NWs, using a heater shaped in the form of a metallic diffusive constriction~\cite{0268-1242-31-10-104001}.

\begin{figure}[t]
	\centering 
	\includegraphics[width=\textwidth]{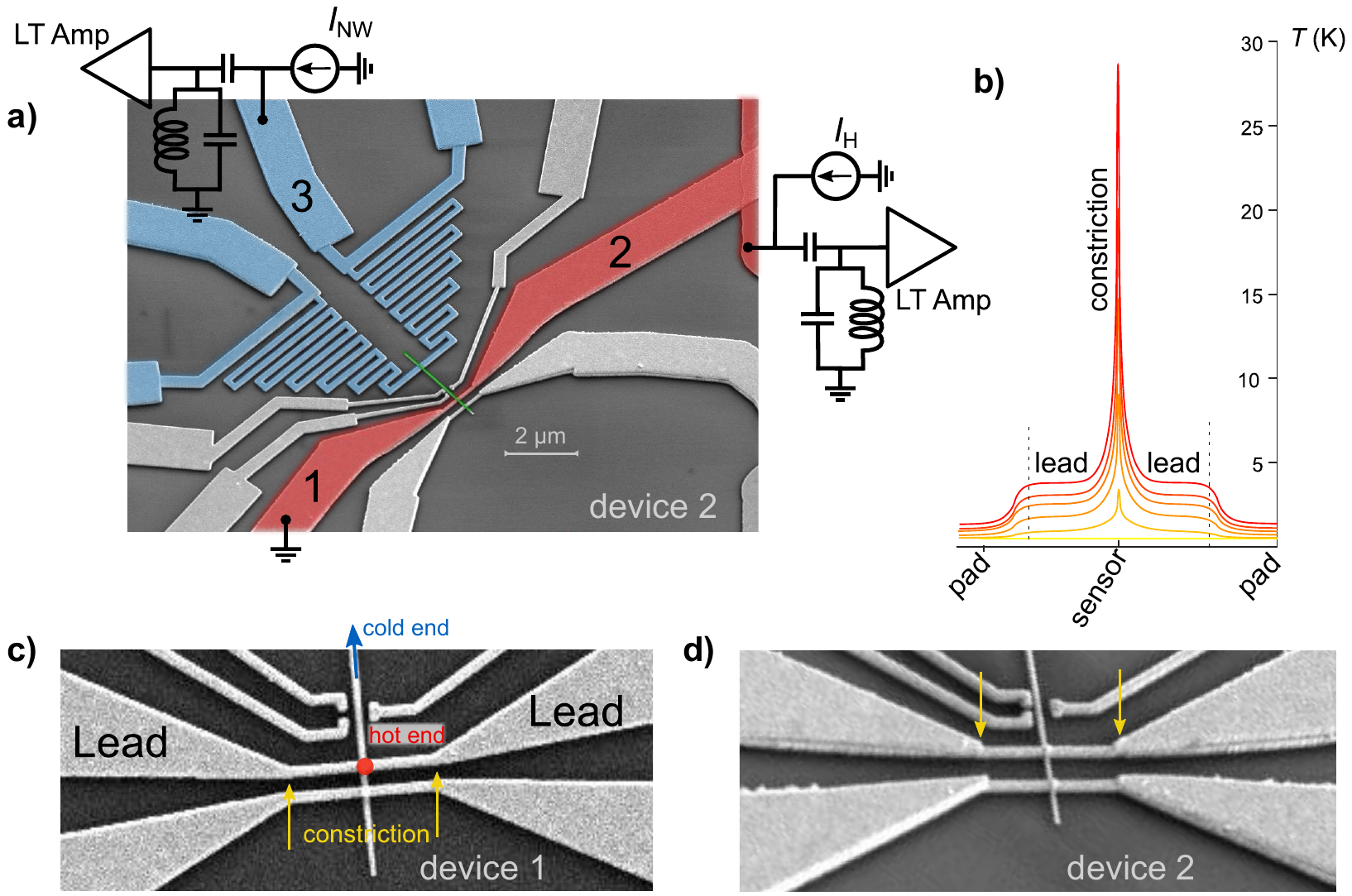}
	\caption{(a\remove{,b}) SEM image\remove{s} of the studied \remove{devices: device 1 (D1, (a)) and} device 2 (D2\remove{, (b)}). The single InAs nanowire (NW, green) on $\rm{SiO_{2}/Si}$ substrate with Ni/Au nanostructure (gray, red, blue) deposited over it. The reddish strip between terminals 1 and 2 is the contact heater device under consideration. Terminal 2 is used to drive the heating current $I_\mathrm{H}$ and to measure the noise from the contact heater, while terminal 1 is connected to the common ground. Terminal 3 is connected to the NW through a meander-shaped strip (blue) and used to drive the current $I_\mathrm{NW}$ and measure the local noise. The shape of the blue strip is not intended to have an impact on the discussed experiments. The three side gates and all other unlabeled contacts were kept unconnected. \change{(b) Numerically simulated spatial temperature profiles along the heater in D1 at different $|I_\mathrm{H}|=2\mu A,\; 1.4\mu A,\; 1.0\mu A,\; 0.6\mu A,\; 0.3\mu A ,\; 0\mu A$.} (c,d) The magnified middle part of the heaters in D1 (c) and D2 (d). Short ($\sim2{\mu}m$)  Ni/Au constriction is connected to huge $\sim{160}\,{\mu}m$ leads. Current $I_\mathrm{H}$  in the constriction heats up the 'hot end' of the NW, while the 'cold end' is kept thermalized at $T_\mathrm{bath}$. The arrows mark the difference {\color{black} in the leads thickness} between D1 and D2 (see text). \remove{(e) Numerically simulated spatial temperature profiles {\color{black} along} the contact heater in D1 at different $|I_\mathrm{H}|=2\mu A,\; 1.4\mu A,\; 1.0\mu A,\; 0.6\mu A,\; 0.3\mu A ,\; 0\mu A$}}. 
	\label{fig1}
\end{figure}    

The question we answer in this work, is whether the accurate knowledge of the thermal bias across a nanostructure (the InAs NW on a $\rm SiO_2$ substrate in our case) is possible without technical challenges imposed by the noise thermometry \change{or other specialized nanoscale sensing techniques.} We will show, that the universal expression for electron temperature in the center of a diffusive heater of resistance $r_{\rm H}$ at a given bath temperature $T_{\rm bath}$~\cite{PhysRevLett.76.3806, PhysRevB.52.4740}:

\begin{equation}
    T=\sqrt{T_{\rm bath}^{2}+\frac{(r_{\rm H}I_{\rm H})^2}{4\mathcal{L}}},
    \label{Temp}
\end{equation}
where $\mathcal{L}=\pi^2k_{\rm B}^2/3e^2$ is the Lorenz number, can be used to calibrate the thermal bias created by the constriction, without performing the noise measurements. A well-known issue in this case is the Joule heating of the metallic leads connecting the constriction to the external current source~\cite{PhysRevB.59.2871}. Using the local and average noise thermometry, we demonstrate a decisive role played by the  non-ideal leads, which can result in a factor of two higher effective heater resistance $r_{\rm H}$ in a typical design. Based on our theoretical analysis, we also propose a \remove{straightforward} strategy to in-situ calibrate the contribution of the leads. \change{The studied effects are based on a well-established physics yet, to our best knowledge, the biasing strategy and analysis method here described have not been reported in the literature.}\remove{ Potentially, this} \change{Our} strategy allows one to determine the thermal bias applied to a device based solely on the knowledge of the heater resistance and, thus, appears to be commonly accessible to experimentalists working with thermal management at the nanoscale.

\section*{\color{black} Devices}

In our experiment we use two types of devices of identical planar architecture shown in the SEM \change{image of Fig.~\ref{fig1}a}\remove{ images reported in Fig.~\ref{fig1}a and~Fig.~\ref{fig1}b and marked as device {1} (D{1}) and device {2} (D{2}), respectively}. Single InAs nanowire\remove{ (see fabrication details in {\bf "Materials and Methods"})}, emphasized with green color, was deposited on the top of $\rm SiO_{2}/Si$ substrate. Colored with light-grey, are $\rm Ni/\color{black} Au$ bilayers deposited by means of e-beam lithography on the substrate, which form ohmic contacts and side gates to the NW (the latter weren't used throughout this work). In our measurements only the terminals numbered 1-3 were used while the others were floating. Reddish  metallic strip between terminals 1 and 2, biased with current $I_{\rm H}$,  serves as the contact heater to the NW. The terminal 2 is connected to the DC external circuit and to the low-temperature amplifier for the average noise measurements, while the terminal 1 is kept grounded (see Fig.~\ref{fig1}\change{a}). On the opposite side, each  NW is connected via the terminal 3 and the bluish meander-shaped strip to the DC measurement circuit and another low-temperature amplifier, in this case for the local noise measurements~\cite{Tikhonov2016}. 

\begin{figure}[t]
	\centering 
	\includegraphics[width=0.8\textwidth]{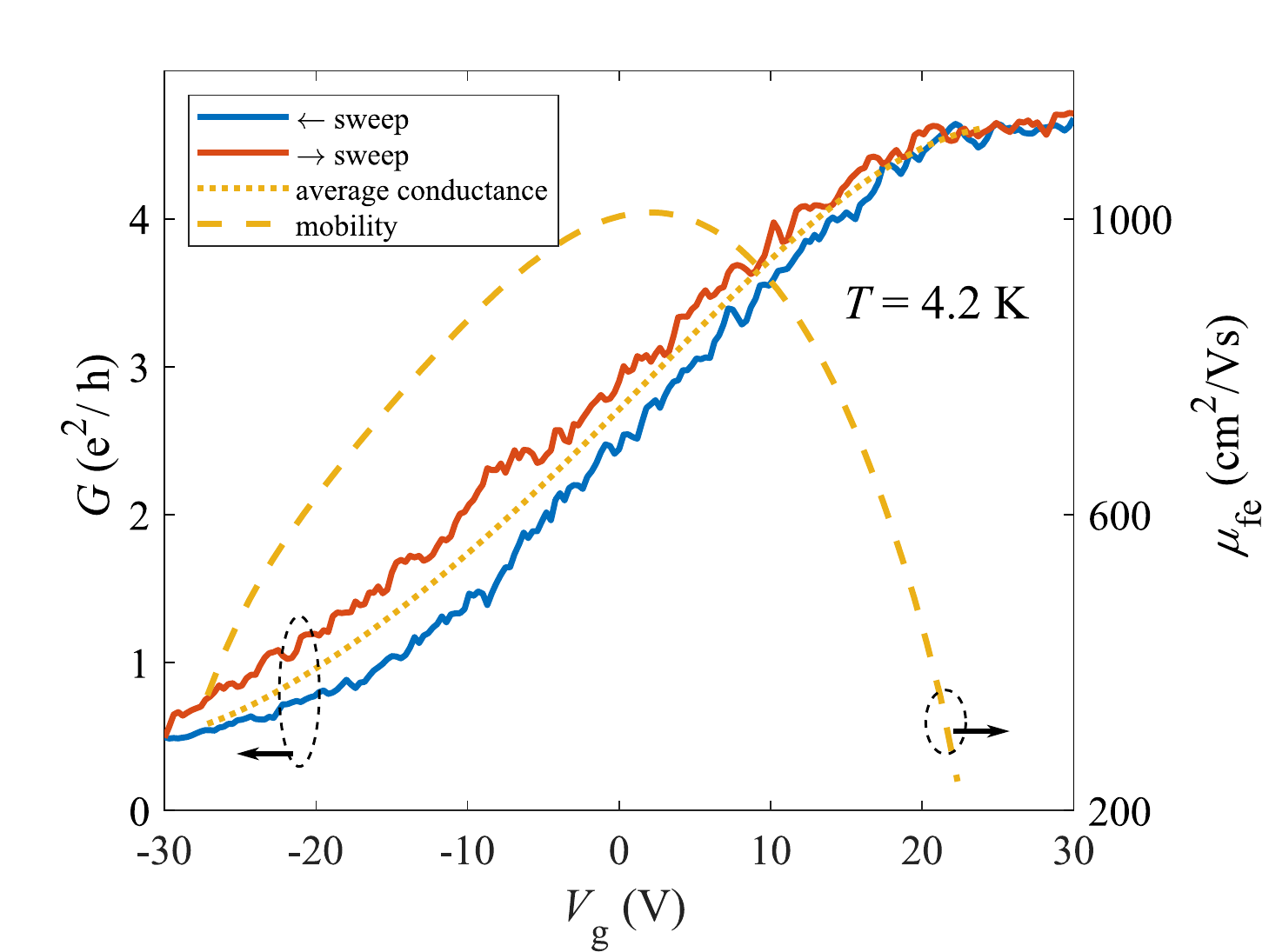}
	\caption{\change{Low temperature transport characterization of InAs NW material. (Left axis) Gate voltage traces of NW conductance at 4.2 K in linear response, measured in a device nominally identical to those used in the experiment. Two solid lines correspond to different sweep directions, see legend. The dotted line is the average curve, which marks average gate voltage position for each value of $G$ calculated from spline-interpolated experimental traces. (Right axis) Field-effect mobility extracted from the gate voltage averaged transconductance is shown by the dashed line, see text.}}. 
	\label{fig_mu}
\end{figure}

\change{In this work we used devices based on InAs NWs, see fabrication details in {\bf "Materials and Methods"}. The high aspect-ratio, the nanometric size and the peculiar electronic transport properties, well known to be diffusive and elastic (energy conserving), make InAs NWs the proper material choice for present experiment. Moreover, the NW resistance falls in a few kOhm range, which is well above the resistances of the connecting metallic terminals and makes InAs a perfect local non-equilibrium sensor~\cite{Tikhonov2016,0268-1242-31-10-104001}. In our experience, these properties are very generic among various types of the InAs NWs, from strongly n-type doped wires grown by chemical beam epitaxy (used in this work) to the undoped catalyst-free wires grown in two different molecular beam epitaxy machines. Fig.~\ref{fig_mu} shows the results of low temperature transport characterization of our InAs NWs. In a separate device, nominally identical to those used throughout this work, we estimated transconductance~\cite{Cui2000} and field effect mobility using the approximate analytical model of Ref.~\cite{Ford2009}. The maximum mobility attained around zero back-gate voltage is $\mu_{\rm fe}\approx 10^3 \mathrm{cm^2/Vs}$ and the corresponding carrier density and mean-free path are about $2\times10^8 \mathrm{cm^{-3}}$ and $30\,$ nm, respectively. The obtained $\mu_{\rm fe}$ is very similar to defect-free doped NWs grown by the same method~\cite{Iorio2018} and is roughly an order of magnitude smaller as compared to the undoped and defect-free InAs NWs~\cite{Schroer2010}. In addition, the gate voltage traces of Fig.~\ref{fig_mu} do not exhibit Coulomb blockade-like conductance oscillations usually associated with crystal defects in InAs~\cite{Schroer2010,Nilsson2016}. This suggests that $\mu_{\rm fe}$ in our devices is likely limited by the ionized impurity or surface scattering. Still we stress again, that the strength of disorder scattering and the value of the carrier density, are not critical parameters for the performance of InAs NWs as local non-equilibrium sensors in present experiment, unlike the elastic transport mechanism.}

Important part of our devices is the constriction in the {\color{black} middle} of the heater strip, see  magnified SEM images in Figs.~\ref{fig1}c and~\ref{fig1}d.  The constriction is represented by a 2\,$\mu$m long and narrow metallic wire, which smoothly evolves into the wide and macroscopically long ($\sim160\,\mu$m) leads. Figs.~\ref{fig1}c and~\ref{fig1}d reveal a crucial difference between the two devices: D{1} was passed through one-step lithography and, thus, a single 120\,nm/10\,nm thick $\rm Ni/Au$ layer appears, while two-step lithography for the  D{2} and another, nominally identical device D{3}, provided a twice thicker metal in the leads area. This difference appears as an abrupt change in evaporated metal thickness marked by yellow arrows in Fig.~\ref{fig1}d. Within each heater strip, the constriction serves as the main heater, whereas the remaining leads represent a non-ideal thermal reservoir,  that is the reservoir with a finite resistivity and thermal conductivity. An example of a numerically simulated temperature profile along the heater strip is shown in Fig.~\ref{fig1}\change{b} for different currents $I_{\rm H}$, see section~\textbf{\nameref{section4}} for the details. Note that the maximum temperature is achieved in the center of the constriction, where the InAs NW connects to the heater via the ohmic contact of negligible electric and thermal resistance ($\sim100\,\Omega$ vs a few k$\Omega$ resistance of the NW). 

In the following, we mostly concentrate on the measurements performed in the device D{1}, except for the measurement of the non-equilibrium electronic energy distribution in the device D{3} in section~\textbf{\nameref{section3}} and the experiments reported in section~\textbf{\nameref{section5}}, where we analyze the role of the leads thickness using the devices D{1} and D{2}. \change{Finally, in the last section we present data from a device of different type, based on Aluminum cross with a tunnel junction in the middle, which is described separately in the text.}

\section*{InAs NW as an energy preserving sensor}\label{section3}

\begin{figure}[t]
	\centering
	\includegraphics[width=\textwidth]{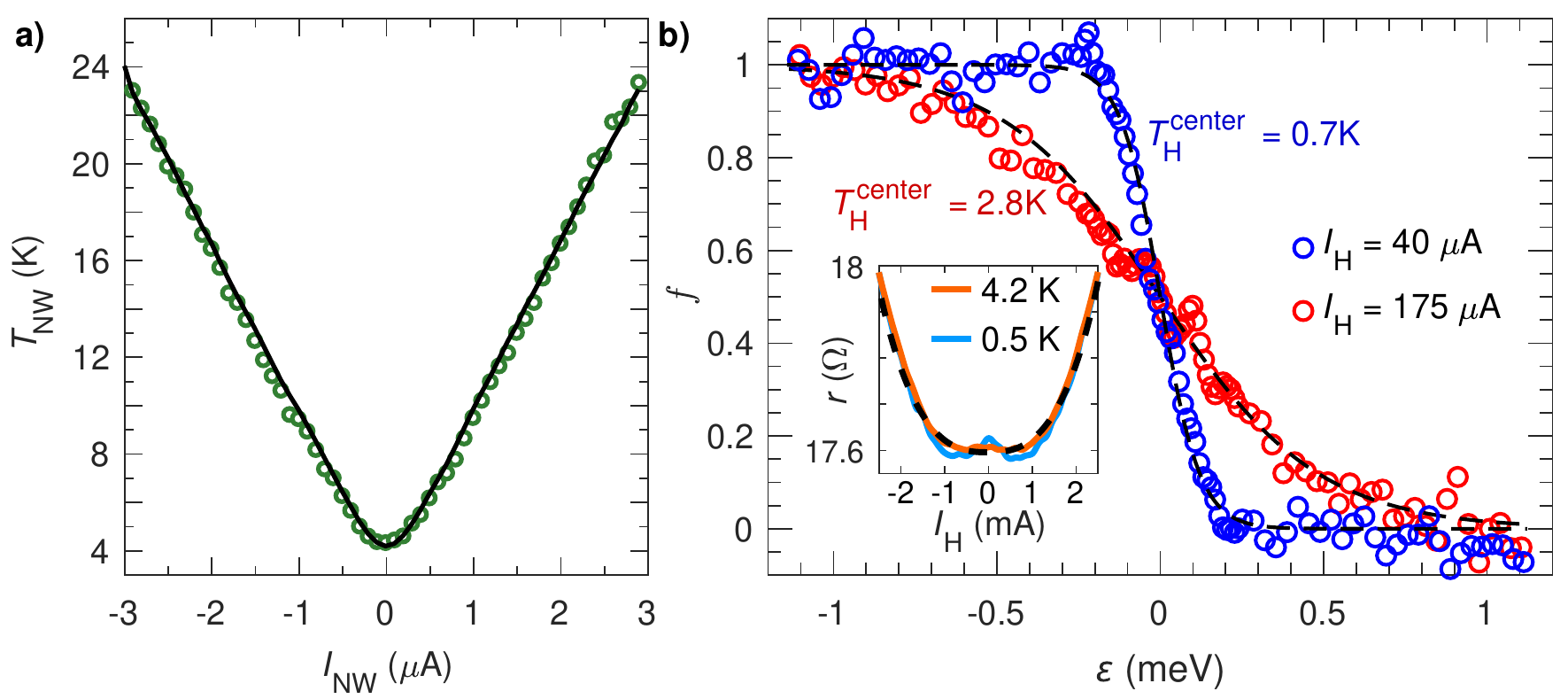}
	\caption{(a) The noise temperature of the NW as a function of $I_\mathrm{NW}$ at $I_\mathrm{H}=0$ and $T_\mathrm{bath} = 4.2$\,K (green symbols). Solid line is the shot-noise prediction with $F = 0.32$, which is close to the universal value for elastic diffusive conductor. (b) The electron energy distribution function in the center of metallic heater, measured via the shot-noise spectroscopy, {\color{black} see Ref.~\cite{7985929} for details}. The blue symbols are extracted from measurement at $I_\mathrm{H} = 40\,\mu$A, the red symbols from measurement at $I_\mathrm{H} = 175\,\mu$A. Dashed lines are equilibrium Fermi-Dirac distribution functions with specified temperatures $T_\mathrm{H}^\mathrm{center}$. The inset shows the nonlinear differential resistance $r$ of the heater between contacts 1 and 2 as a function of $I_\mathrm{H}$ at $T_\mathrm{bath}=4.2$\,K (solid orange line) and $T_\mathrm{bath}=0.5$\,K (solid blue line) in device D1. The dashed line is a numerical  prediction assuming a power-law temperature dependent correction to the conductivity (see text). }
	\label{fig2}
\end{figure}

The goal of this section is to summarize the capabilities of a diffusive InAs NW as a sensor of the local temperature~\cite{Tikhonov2016} and the local energy distribution~\cite{7985929}. This is the only experimental section in which the NW is connected to the external current source and biased with the current $I_{\rm NW}$, which flows between the terminals 3 and 1 in Fig.~\ref{fig1}\change{a}. We start from the characterization of the transport regime in the InAs NW measuring its shot noise at a bath temperature $T_{\rm bath}=4.2\,$K. In Fig.~\ref{fig2}a we plot the measured NW noise temperature $T_{\rm NW}=S_{\rm I} R_{\rm NW}/4k_{\rm B}$, where $S_{\rm I}$ is the noise spectral density and $R_{\rm NW}$ is the differential resistance of the NW  as a function of $I_{\rm NW}$,  while $I_{\rm H}=0$.

The crossover from the equilibrium Johnson-Nyquist noise $S_{\rm I}=4kT_{\rm bath}/R_{\rm NW}$ at $I_{\rm NW}=0$ to linear current dependence $S_{\rm I}=2eFI_{\rm NW}$, where $F$ is a Fano factor, is observed and persists up to $T_{\rm NW}\approx24$\,K (symbols). The theoretical fit (solid line) meets experimental data at $F=0.32$ which is very close to the universal value for diffusive conductors without electron-phonon relaxation $F=1/3$~\cite{NAGAEV1992103,PhysRevB.46.1889}. Thus, a quasiparticle energy is preserved along the NW, making it ideally suited for local noise sensing~\cite{Tikhonov2016}.

As usual in the case of elastic diffusion~\cite{NAGAEV1992103}, the electronic energy distribution (EED) at a given location $x$ along the NW (in units of its length), $f_\varepsilon(x)$, obeys the Laplace's equation ${\partial^2 f_\varepsilon(x)}/{\partial x^2}=0$. The solution is a linear combination of the EEDs $f_\varepsilon(0)$ and $f_\varepsilon(1)$ given by the external boundary conditions at the two NW ends:

\begin{equation}
   f_\varepsilon(x)=\bigg(1-x\bigg)f_\varepsilon(0)+xf_\varepsilon(\remove{L}\change{1})
    \label{eq3}
\end{equation}

 The cold end of the InAs NW (connected to the terminal 3 at $x=0$) is always kept in equilibrium with the corresponding EED $f_\varepsilon(0)=(1+\exp(-\varepsilon/k_{\rm B}T_{\rm bath}))^{-1}$. In the following we focus on the experiments with a finite heater current $I_{\rm H}$, thereby the second boundary condition in the eq.~(\ref{eq3}) is non-equilibrium. We evaluate the corresponding EED at $x=1$ by utilizing the energy resolved local noise spectroscopy~\cite{Gramespacher1999,tikhonov2020energy}. In this experiment, performed with a separate device nominally identical to D{2} at $T_{\rm bath}\approx100$\,mK, both bias currents $I_{\rm NW}$ and $I_{\rm H}$ are finite. Also the $I_{\rm H}$ was chosen high enough, to avoid a problem with the noise analysis caused by a non-linearity of the NW resistance, see Ref.~\cite{7985929} for details. In Fig.~\ref{fig2}b the measured $f_\varepsilon(1)$ is shown for two different $I_{\rm H}$ values along with the corresponding fits at the same bath temperature (symbols and dashed lines, respectively). The measured $f_\varepsilon(1)$ is indistinguishable from the Fermi-Dirac EED $(1+\exp(-\varepsilon/k_{\rm B}T_{\rm H}^{\rm center}))^{-1}$, where $T_{\rm H}^{\rm center}$ is the temperature of the NW's hot end,  which coincides with the local temperature in the center of the metallic heater constriction. Note that as soon as the EEDs at the NW's ends are known, there is a unique correspondence between the measured $T_{\rm NW}$ and the $T_{\rm H}^{\rm center}$, since $T_{\rm NW}=\int dx\int f_\varepsilon(1-f_\varepsilon)d\varepsilon$, see Ref.~\cite{NAGAEV1992103}. 

The observation of the locally equilibrium EEDs in Fig.~\ref{fig2}b implies strong thermalization of the charge carriers in the metallic heater constriction, even at relatively low temperatures of $T_{\rm bath}\approx100$\,mK  and $T_{\rm H}^{\rm center}\approx0.7$\,K. This is in contrast to a naive expectation of a double-step EED generated by the current $I_{\rm H}$~\cite{PhysRevLett.79.3490}. We believe that the reason for such a strong thermalization is the electron-electron collisions in presence of a spin-flip scattering~\cite{PhysRevLett.86.2400,PhysRevLett.90.076806,PhysRevB.64.033301,tikhonov2020energy}. In our case such scattering is inevitable owing to the ferromagnetic Ni layer used in metallization. An independent signature of the spin-flip scattering comes from a zero-bias Kondo-like peak observed in a non-linear heater resistance in dependence of $I_{\rm H}$ at $T_{\rm bath}=0.5$\,K, see the blue curve in \change{the} inset of Fig.\ref{fig2}b.

\section*{Local vs average heater thermometry: impact of the non-ideal leads}\label{section4}

\begin{figure}[t]
	\centering
	\includegraphics[width=\textwidth]{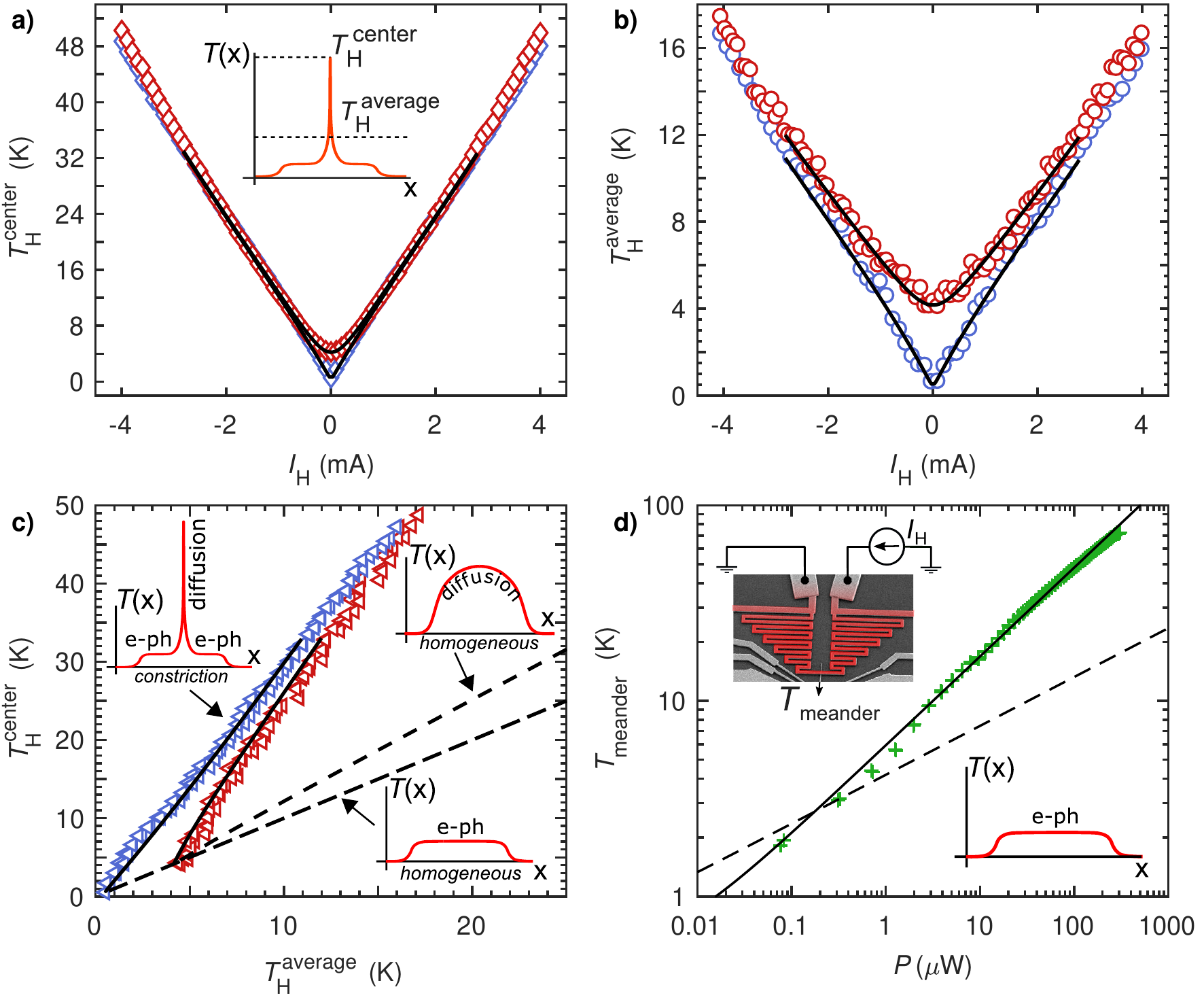}
	\vspace{-0.8cm}
	\caption{Local and average noise thermometry applied to the characterization of the temperature profile in the contact heater. (a) The measured local temperature in the center of Ni/Au constriction $T\mathrm{_H^{center}}$ at $T_{\rm bath}=0.5$\,K (blue diamonds) and $4.2$\,K (red diamonds) along with the corresponding numerical simulations (black lines). (b) The same as (a) but for average temperature $T\mathrm{_H^{average}}$ of the heater. The inset on (a) shows an example of simulated spatial temperature profile along the heater $T(x)$ with marked $T\mathrm{_H^{center}}$ and $T\mathrm{_H^{average}}$ at $I_{\rm H}=2.5$\,mA. (c) $T\mathrm{_H^{center}}$ plotted as a function of $T\mathrm{_H^{average}}$. The experimental data from (a) and (b) are combined and shown by blue symbols ($T_{\rm bath}=0.5$\,K) and red symbols ($T_{\rm bath}=4.2$\,K). The solid lines are the result of numerical calculations based on heat balance equation (see text) for both bath temperatures. The dashed lines for $T_{\rm bath}=0.5$\,K are predictions for the two limiting cases of the diffusion cooling and the electron-phonon cooling in a homogeneous conductor. The corresponding spatial temperature profiles $T(x)$ are shown in the nearby insets. (d) Measured local temperature in the center of a long meander vs dissipated power at $T_{\rm bath}=0.5$\,K. Green crosses correspond the experimental data, while the solid line shows the best fit for the power law e-ph cooling $P_{\rm eph}=\Sigma_{\rm eph}(T^m-T_{\rm bath}^m)$ with $m=2.2$. The dashed line shows the estimate for Kapitza resistance heat flow bottleneck $A\Sigma_{\rm K}(T^{4}-T_{\rm bath}^{4})$ with the parameters given in the text. The insets demonstrate spatial temperature profile $T(x)$ along the meander and its microscopic image.
	}
	\label{fig3}
\end{figure}

In this section we discuss the impact of the non-ideal leads of the metallic heater constriction on the thermal biasing. Here we supplement the local noise thermometry~\cite{Tikhonov2016}, which provides the knowledge of the temperature $T_{\rm H}^{\rm center}$ at the NW contact position, i.e. at the center of the metallic constriction, with the conventional average noise thermometry~\cite{PhysRevLett.76.3806,PhysRevB.59.2871,PhysRevLett.81.2982,Henny1997,Larocque2020}. The latter approach is utilized via a measurement of the current noise  of the heater strip as a whole, which is picked-up by the low-temperature amplifier at terminal 2. The measured signal in this case is the average noise temperature $T_{\rm H}^{\rm average}$, which is given by the average of position-dependent heater temperature with the weight of local Joule heat~\cite{PhysRevB.96.245417}. The difference between the $T_{\rm H}^{\rm center}$ and $T_{\rm H}^{\rm average}$, along with the spatial temperature distribution in the heater at $T_{\rm bath}=0.5\,$K and $I_{\rm H}=2.5\,$mA, is demonstrated in the inset of Fig.~\ref{fig3}a. In \change{the} body of Fig.~\ref{fig3}a we plot the measured temperature $T_{\rm H}^{\rm center}$ in the center of the metallic constriction against the bias current $I_{\rm H}$ at $T_{\rm bath}=0.5$\,K and $T_{\rm bath}=4.2$\,K (symbols). At $I_{\rm H}=0$, obviously, $T_{\rm H}^{\rm center}=T_{\rm bath}$, while at increasing $|I_{\rm H}|$ the measured temperature passes to the linear dependence up to $T_{\rm H}^{\rm center}\sim50$\,K. As shown in Fig.~\ref{fig3}b, the heater-averaged temperature $T_{\rm H}^{\rm average}$ behaves similarly, also demonstrating a linear dependence  on the bias current at large enough $I_{\rm H}$ (symbols).  In this case, however, the measured temperature increase is considerably smaller. Such a strong discrepancy between the local and average temperatures highlights the main feature of our heater geometry, which is designed as a macroscopic metallic strip with a short constriction, see Fig.~\ref{fig1}. 

In Fig.\ref{fig3}c we plot the $T_{\rm H}^{\rm center}$ in dependence of the $T_{\rm H}^{\rm average}$ (symbols), and compare the experimental results with the two limiting cases expected in a homogeneous conductor without constriction (dashed lines). Although the spatial temperature profiles are different in the case of a diffusion cooled conductor, sketched next to the upper dashed line, and in the case of the electron-phonon (e-ph) cooled conductor, sketched next to the lower dashed line, in both cases one obtains $T_{\rm H}^{\rm center}/T_{\rm H}^{\rm average}\approx 1$. By contrast, in our experiment $T_{\rm H}^{\rm center}/T_{\rm H}^{\rm average}\approx3$, which is a direct consequence of our heater design. As we discuss below, the constriction and the leads in our case are in the regimes of diffusion cooling and e-ph cooling, respectively, the corresponding temperature profiles are sketched in Fig.~\ref{fig3}c.

The solid lines in Figs.~\ref{fig3}a,~\ref{fig3}b and~\ref{fig3}c represent the results of numerical calculations used to fit the experimental data and characterize the parameters of the e-ph cooling in our devices. The underlying physics is captured by the heat balance equation:
\begin{equation}
	-\nabla\left(\kappa_e\nabla T\right)=\change{ \frac{j_{\rm H}^2}{\sigma_{\rm H}} }-P_{\rm eph}, \label{eq_balance}
\end{equation}
where $\kappa_e=\sigma_{\rm H}\mathcal{L}T$ is the Wiedemann-Franz heat conduction of the electronic system, which is responsible for the diffusion cooling mechanism in the heater \change{and $j_{\rm H}$ is the local current density in the heater}. The first term on the rhs of the eq.~(\ref{eq_balance}) accounts for the Joule heat production in the heater, whereas the second term stands for the e-ph cooling, both per unit volume. The best fits, capable to explain the data of in Figs.~\ref{fig3}a,~\ref{fig3}b and~\ref{fig3}c up to $|I_{\rm H}|=2.5$\,mA are obtained with the following parameters. First, we assumed the power-law e-ph cooling  $P_{\rm eph}=\Sigma_{\rm eph}(T^m-T_{\rm bath}^m)$ with $\Sigma_{\rm eph}=6\times10^9\,\rm W/m^3K^{2.5}$ and the exponent of $m=2.5$. This unusual exponent both provides the best fit to the experimental data in constrictions and is consistent with an independent e-ph cooling measurement in a homogeneous heater, as described below. Second, we took the $T$-dependence of the heater conductivity into account via $\sigma_{\rm H}=\sigma_{\rm H}(T_{\rm bath})-\alpha(T^{3}-T_{\rm bath}^{3})$, where $\sigma_{\rm H}(T_{\rm bath})=6.3\times10^7\,\rm S/m$ is measured at $I_{\rm H}=0$ and $T$ is the position-dependent electronic temperature in the heater. The value of the parameter $\alpha=90\,\rm S/mK^{3}$ is consistent with the observed dependence of the total heater resistance at $T_{\rm bath}=4.2\,$K, see the inset of Fig.~\ref{fig2}b,  and captures the main trend at $T_{\rm bath}=0.5\,$K.

An independent study of the e-ph cooling power $\Sigma_{\rm eph}$ is achieved via a measurement of the local temperature in the center of a long meander-shaped heater, depicted in the inset of Fig.~\ref{fig3}d. We choose the meander for its length of $\approx 105\,\rm\mu$m, which is significantly longer than e-ph relaxation length (see below). Hence, we expect a flat temperature profile along the meander, such that the measured local temperature is the same as the electron temperature everywhere else in the meander. Fig.~\ref{fig3}d shows the measured temperature vs dissipated power at $T_{\rm bath}=0.5$\,K. The data (symbols) crosses to the power-law behavior, which is best described by $m\approx2.2$ and $\Sigma_{\rm eph}\approx 5.9\times10^9\,\rm W/m^3K^{2.2}$ (solid line), which is pretty close to the value of $\Sigma_{\rm eph}$ used in our numerical calculations. Again, the exponent of $m\approx2.2$ is unusual and is much smaller than the conventional value of $m\approx5$ observed on a sapphire~\cite{Roukes1985} and oxidized silicon~\cite{Wang2018} substrates at sub-1K temperatures. This indicates that the e-ph cooling at much higher $T$ used in our experiment is considerably bottlenecked by the $\rm SiO_2$ substrate. This bottleneck is very different from the  Kapitza resistance, which is caused by the acoustic mismatch at the interface between the metal film and the insulating substrate. The corresponding heat outflow of $A\Sigma_{\rm K}(T^{4}-T_{\rm bath}^{4})$, where $A=26\,\rm\mu m^2$ is the area covered by the meander and $\Sigma_{\rm K}=125\, \rm pW/\mu m^2K^{4}$ is taken from Ref.~\cite{Roukes1985}, is shown by the dashed line in Fig.~\ref{fig3}d. We observe that  for $T\gtrsim5\,$K the experimental data lies substantially above the dashed line, illustrating a different and much stronger bottleneck mechanism functioning in this temperature range. On the one hand, this is not too surprising, given the fact~\cite{Zeller1971} that the phonon mean-free path in amorphous $\rm SiO_2$ rapidly decays as $\propto T^{-4}$ at increasing temperature above 5\,K. On the other hand, the presence of such a non-universal bottleneck strongly complicates numerical modeling, emphasizing the need for accurate experimental calibration of the thermal bias in the experiments.

  We conclude this section with an estimation of the e-ph relaxation length, which is defined as follows. In a uniform conductor at $I_{\rm H}=0$ a small temperature difference $\delta T\ll T_{\rm bath}$ decays exponentially in space. In this case the solution of the eq.~(\ref{eq_balance}) is $\delta T\propto\exp(-x/l_{\rm eph})$, where $l_{\rm eph}=\sqrt{\sigma_{\rm H}\mathcal{L}/m\Sigma_{\rm eph}T^{m-2}}$ is the e-ph relaxation length. Using the fit parameters of Figs.~\ref{fig3}a,~\ref{fig3}b and~\ref{fig3}c we obtain $l_{\rm eph}\approx7\,\mu$m and $l_{\rm eph}\approx12\,\mu$m, respectively, at $T=4.2\,$K and $T=0.5\,$K in our Ni/Au metallic heaters. This estimate is in very good agreement with the above conclusion that the constriction and the leads of our heaters are, respectively, much shorter and much longer compared to the $l_{\rm eph}$.

\section*{Quantitative strategy for thermal biasing in linear response}\label{section5}

{\color{black}

In Fig.~\ref{fig4}a  we plot the temperature in the center of the heater constriction measured via local noise thermometry in dependence of $I_{\rm H}$ in two devices at $T_{\rm bath}=4.2\,$K (symbols). The data clearly capture a systematic effect of the thickness of the metallic heater leads, which we express in terms of the number $n$ of the 120\,nm/10\,nm thick Ni/Au bilayers. The $T_{\rm H}^{\rm center}$ is considerably reduced in D2 ($n=2$) as compared to the case of D1 ($n=1$), which is perfectly consistent with the results of the numerical calculation of the eq.~(\ref{eq_balance}) (dashed, dash-dotted and thin solid lines). The thick solid line in Fig.~\ref{fig4}a also shows the calculation in the limit of $n\rightarrow\infty$, which corresponds to the idealized situation of the leads with zero electrical and heat resistances.  The data of Fig.~\ref{fig4}a demonstrate that the experimental data for $n=1,\,2$  substantially differ from each other and from the $n\rightarrow\infty$ limit, emphasizing that non-ideal leads have a strong impact on the thermal biasing in reasonable experimental configurations.

\begin{figure}[t]
	\centering
	\includegraphics[width=\textwidth]{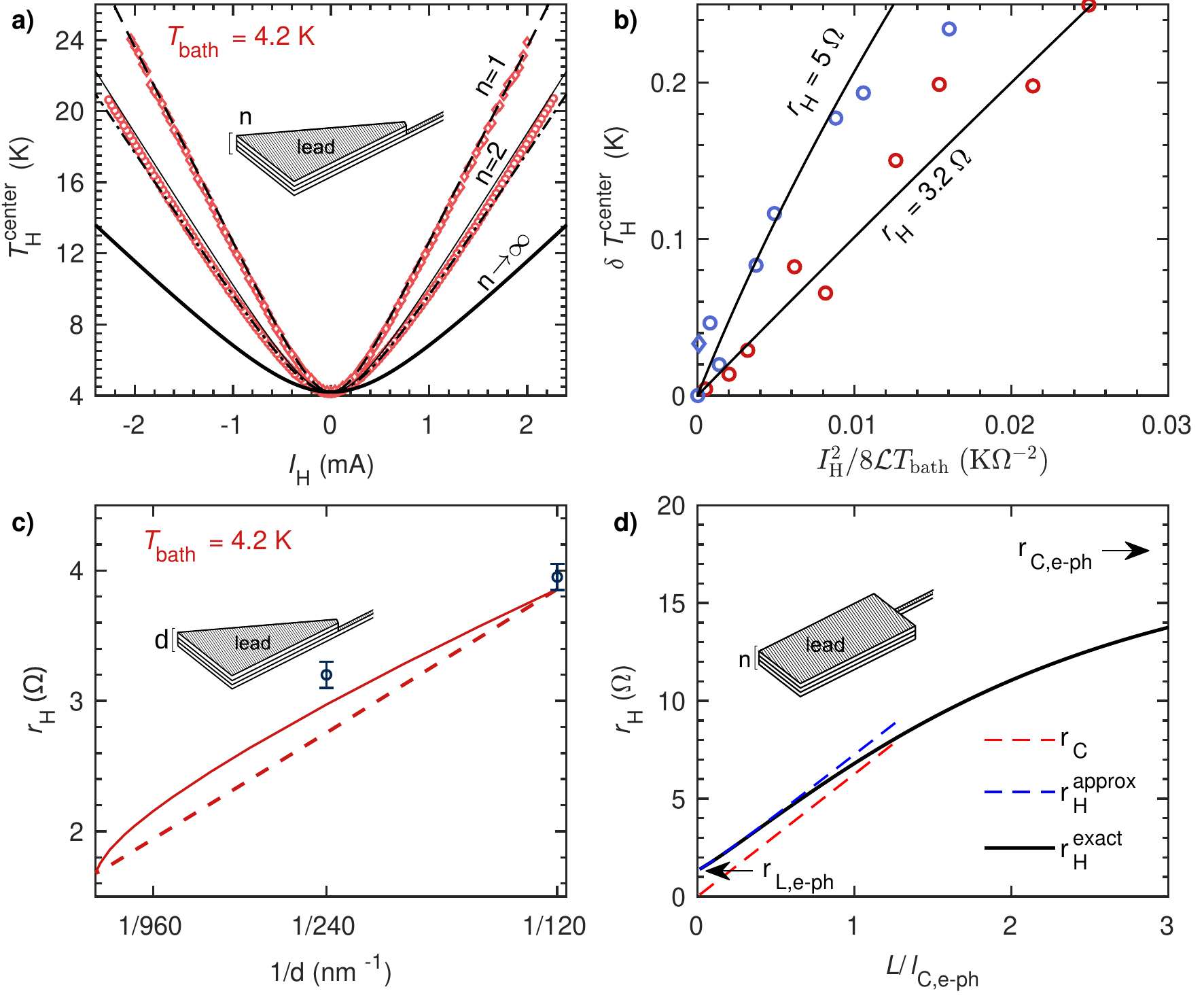}
	\vspace{-0.8cm}
	\caption{(a) The measured temperature in the center of the Ni/Au constriction $T\mathrm{_H^{center}}$ as a function of the heating current $I_\mathrm{H}$ at $T_{\rm bath}=4.2$\,K for samples with the leads composed of $n=1$ and $n=2$ layers (red symbols, see text). The dashed line is the numerical calculation for $n=1$. The dashed-dot line and thin solid line are numerical calculations for $n=2$ with volume and surface e-ph heat outflow respectively. The thick solid line is calculated for the case of ideal leads ($n\to\infty$). The inset depicts the schematic of the trapezoidal leads used in the experiment. (b) The similar measurements as (a) in a form of temperature increment $\delta T\mathrm{_H^{center}}=T\mathrm{_H^{center}}-T_{\rm bath}$ at $T_{\rm bath}=0.5$\,K (blue circles) and $4.2$\,K (red circles) for $n=2$ as a function of a normalized squared heating current $I_{\rm H}^2/8\mathcal{L}T_{\rm bath}$. The lines are the corresponding fits with eq.~(\ref{Temp}) with  $r\mathrm{_H}$ increasing at decreasing $T_{\rm bath}$. (c) The experimental (symbols) and numerically calculated (lines) values of $r\mathrm{_H}$ in the limit $\delta T\ll T_{\rm bath}$ plotted as a function of the inverted thickness of leads $1/d$ at $4.2$\,K. The calculations are built for the volume (dashed line) and the surface (solid line) e-ph heat outflow. Note, that all theoretical predictions meet at $r\mathrm{_H}=r_\mathrm{C}$ when $d\to\infty$, where $r_\mathrm{C}$ is the resistance of the heater constriction. (d) Strategy: effective heater resistance vs the constriction length, normalized by e-ph relaxation length $L/l_{\rm C,e-ph}$ (see text). The solid line shows the exact calculation while the upper dashed line corresponds the approximation of eq.~(\ref{solution_short}). The lower dashed line shows the constriction resistance $r_{\rm C}$. Arrows on the lhs and rhs mark the limiting values of the $r_\mathrm{H}$, respectively, for $L=0$ and $L\rightarrow\infty$.
	}
	\label{fig4}
\end{figure}

Below we focus on the issue of thermal biasing in the linear response regime, that is in the limit of $\delta T\ll T_{\rm bath}$, where the eq.~(\ref{eq_balance}) can be solved analytically. Inspired by the eq.~(\ref{Temp}), we express the temperature rise in the center of a constriction with a cross-section $A_{\rm C}$ connected to the leads of a uniform cross-section $A_{\rm L}$ (see the sketch in Fig.~\ref{fig4}d) as follows:

\begin{equation}
	\delta T_{\rm H}^{\rm center} \equiv T_{\rm H}^{\rm center}-T_{\rm bath} = \frac{\left(r_{\rm H}I_{\rm H}\right)^2}{8\mathcal{L}T_{\rm bath}} \label{solution_dT} 
	\end{equation}
\begin{equation}
r_{\rm H}^2  = \left(1-\alpha\right)r_{\rm C, e-ph}^2+\alpha r_{\rm L, e-ph}^2  	\label{solution_r}
	\end{equation}
\begin{equation}
\alpha = \left[ \frac{r_{\rm L, e-ph}}{r_{\rm C, e-ph}} \sinh{\left(\frac{L}{2l_{\rm C, e-ph}}\right)}+\cosh{\left(\frac{L}{2l_{\rm C, e-ph}}\right)} \right]^{-1} \label{solution_alpha}, 
\end{equation}
where $L$ is the length of the constriction and $r_{\rm C, e-ph}=2\sqrt{2}l_{\rm C, e-ph}/(\sigma_{\rm H}A_{\rm C})$ and $r_{\rm L, e-ph}=2\sqrt{2}l_{\rm L, e-ph}/(\sigma_{\rm H}A_{\rm L})$ are the resistances of a wire with a cross-section of the constriction and the leads, respectively, and the length determined by the corresponding e-ph relaxation length. As follows from the eqs. (\ref{solution_dT}-\ref{solution_alpha}), depending on $L$, the $\delta T_{\rm H}^{\rm center}$ interpolates between the solutions with $\alpha=1$ for $L=0$ and  with $\alpha=0$ for $L\rightarrow\infty$ (see the solid line in Fig.~\ref{fig4}d). Both limiting cases correspond to thermal biasing with a uniform metallic strip, for which the effective heater resistance, $r_{\rm H}$, is given solely by the {\it a-priori} unknown e-ph relaxation length. 

The situation is different when the central part of the heater is shaped as a constriction, short compared with the e-ph relaxation length, $L\ll l_{\rm C, e-ph}$, which is the case in our experiments. Here, $r_{\rm H}$ includes the known electrical resistance of the constriction $r_{\rm C}=L/(\sigma_{\rm H}A_{\rm C})$ and is given by:

\begin{equation}
r_{\rm H}^2  = r_{\rm C}^2+\sqrt{2}r_{\rm C}r_{\rm L, e-ph}+r_{\rm L, e-ph}^2 \label{solution_short}
	\end{equation}

Eq.~(\ref{solution_short}) predicts that the actual value of $r_{\rm H}$ should vary as a function of both $T_{\rm bath}$ and  $A_{\rm L}\propto n$. This is indeed observed in our experiment, as shown in Figs.~\ref{fig4}b and~\ref{fig4}c. In Fig.~\ref{fig4}b, we plot $\delta T_{\rm H}^{\rm center}$ as a function of the quantity $I_{\rm H}^2/8\mathcal{L}T_{\rm bath}$ in device D2 at $T_{\rm bath}=0.5\,$K (blue symbols) and $T_{\rm bath}=4.2\,$K (red symbols)\change{ , along with the fits according to eq.~(\ref{Temp})}. We observe that the \change{initial} linear slope of the data for $\delta T_{\rm H}^{\rm center}\ll T_{\rm bath}$ increases at decreasing temperature. This corresponds to the increase of $r_{\rm H}$ owing to the increased e-ph relaxation length and, thus, of the $r_{\rm L, e-ph}$, see eq.~(\ref{solution_short}). \change{Note the deflection of the blue symbols from the corresponding line at higher currents. Here the measured $\delta T_{\rm H}^{\rm center}$ is not small enough compared to the bath temperature $T_{\rm bath}=0.5\,$K and the lowest order expansion (\ref{solution_dT}) fails. The higher order correction results in a decrease of the e-ph relaxation length, hence the decrease of the $r_{\rm H}$, see Eq.~(\ref{solution_short}), and slowing down of the $\delta T_{\rm H}^{\rm center}$ increase.}

\begin{figure}[t]
	\centering
	\includegraphics[width=\columnwidth]{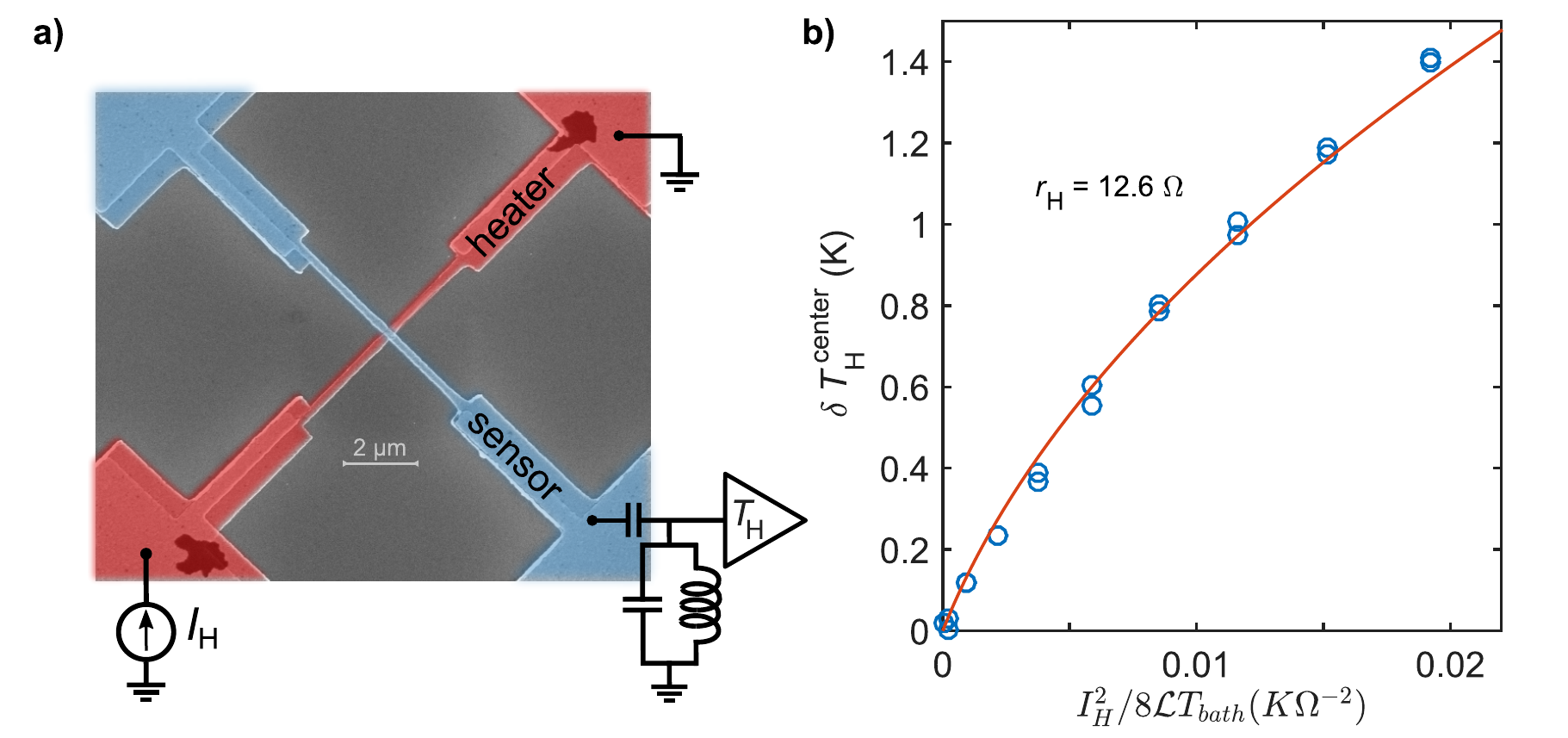}
	\vspace{-0.8cm}
	\caption{\change{(a) SEM image of the Al-based device designed with intentionally small contribution of the lead resistance. The device is represented by a cross with the lower (reddish) Al layer playing role of the heater with a constriction. Pad to pad resistance is about 12\,$\Omega$. This layer is connected to the current source as shown in the schematics. One of the pads of the upper (blueish) Al layer is connected to the noise measurement circuit, whereas the second pad is left floating. In the center of the cross a thin in-situ grown Al-oxide tunnel barrier with a resistance of about 5\,k$\Omega$ is placed, which divides the two Al layers and plays the role of the local noise sensor, instead of the InAs NW in devices of Fig.~\ref{fig1}. For more details on this device see Ref.~\cite{tikhonov2020energy}. (b) Increment of the temperature in the center of the heater constriction plotted as a function of normalized $(I_{\rm H})^2$ at $T_{\rm bath}\approx0.5\,$K (symbols). The fit to eq.~(\ref{Temp}) is shown by the solid line with the heater resistance parameter given in the legend. }	}
	\label{fig5}
\end{figure}

In Fig.~\ref{fig4}c we plot the values of $r_{\rm H}$ in dependence of the inverse thickness of the leads, $d\propto n$, obtained in devices D1 and D2 at $T_{\rm bath}=4.2\,$K (symbols). As seen from the plot, the measured $r_{\rm H}$ can exceed $r_{\rm C}\approx1.8\,\rm\Omega$ by about a factor of 2 or more for the chosen device geometry, demonstrating again that the effect of the non-ideal leads is by no means negligible. A strong difference between the $r_{\rm H}$ and $r_{\rm C}$ in our experiment is not surprising, for the constriction and leads of the heater are made of the same material and the cross-section area of the leads is still not big enough to make the $r_{\rm L, e-ph}\propto 1/A_{\rm L}$ small. A natural way to minimize the spurious effect of the leads is to make them from a material of much higher conductivity and choose $A_{\rm L}$ as big as possible. \remove{However, this task is well known to be difficult to realize experimentally~\cite{PhysRevB.59.2871}.} \change{We explored this possibility with a separate device based on an aluminum cross shown in Fig.~\ref{fig5}a. Two branches of the cross are connected by a thin tunnel barrier, achieved by in-situ oxidation of the Al. The lower (reddish) branch represents a heater with a 3\,$\mu$m long constriction in the middle. The design of the leads was chosen such that $A_{\rm L}$ grows quickly away from the constriction and the total lead resistance is small compared to the $r_{\rm C}$. The upper (blueish) branch connects the tunnel junction, which plays the role of the local noise sensor in this device, to the low temperature amplifier. The measured local temperature in the middle of the constriction is plotted in Fig.~\ref{fig5}b as a function of normalized $(I_{\rm H})^2$ at $T_{\rm bath}\approx0.5\,$K (symbols), along with the best fit to eq.~(\ref{Temp}). Unlike in our "careless" NW-based devices, here the lead resistance has a smaller contribution and fitted $r_{\rm H}=12.6\,\mathrm{\Omega}$ is much closer to the $r_{\rm C}\approx8.3\,\mathrm{\Omega}$, see the legend. Correspondingly, the eq.~(\ref{Temp}) adequately describes the data in a much wider range of $\delta T_{\rm H}^{\rm center}$ as compared to the dataset of Fig.~\ref{fig4}b for similar $T_{\rm bath}$. Moreover, we observed that the device of Fig.~\ref{fig5}a supports elastic diffusive transport regime and at even higher $I_{\rm H}$ the local noise spectroscopy in spirit of Fig.~\ref{fig2}b reveals a non-thermal double-step-shaped energy distribution in the center of the heater, see Ref.~\cite{tikhonov2020energy} for details. Obviously, however, this example is special and in most cases the contribution of the leads cannot be made negligible for technical or other reasons.} In the following we discuss a strategy for the calibration of the thermal bias in experimentally feasible designs \change{of that kind.} \remove{with a non-negligible contribution of the leads.}

One approach is to make use of a variation of the thickness $d$ of the leads in spirit of the present experiment. In this way, using eq.~(\ref{solution_short}) and extrapolating the experimental data (e.g., the TE data) towards $r_{\rm H}(d\rightarrow\infty)=r_{\rm C}$ one can calibrate the thermal bias quantitatively. Unfortunately, the underlying scaling of $r_{\rm L, e-ph}$ with $d$ is {\it a-priori} unknown and can vary between $r_{\rm L, e-ph}\propto d^{-1}$ and $r_{\rm L, e-ph}\propto d^{-1/2}$, respectively, for the e-ph relaxation scaling with the volume and with the surface of the leads. The latter scaling corresponds to a situation when a bottleneck for the e-ph relaxation occurs via a poor heat conduction of the substrate, which is likely the case in present experiment. The dashed and solid lines in Fig.~\ref{fig4}c show the calculated dependencies $r_{\rm H}$ vs $d$ for these two model situations, respectively. In spite of very different functional dependencies, both models are capable to describe the experimental data with roughly the same accuracy, demonstrating how generally challenging it is to use the extrapolation $r_{\rm H}(d\rightarrow\infty)=r_{\rm C}$  for the calibration of thermal bias. Similarly, one can investigate the dependence of $r_{\rm H}$ on the $T_{\rm bath}$, which occurs owing to a variation of the e-ph relaxation via $r_{\rm L, e-ph}\propto l_{\rm L, e-ph}$. This approach is also difficult to realize for the underlying temperature dependence is not {\it a-priori} known.

The most reliable calibration can be achieved by varying $L$, the length of the constriction, which corresponds to a variation of $r_{\rm C}$. In this approach, the contribution $r_{\rm L, e-ph}$ of the non-ideal leads remains constant and the uncertainty associated with the features of the e-ph relaxation in the leads becomes irrelevant. The scaling of the effective heater resistance $r_{\rm H}$ as a function of $L$, in units of the corresponding e-ph relaxation length, is shown in Fig.~\ref{fig4}d. This graph is obtained for the parameters of a device similar to our device D1 ($n=1$) at $T_{\rm bath}=4.2$\,K, but without an intermediate trapezoidal transition region between the constriction and the leads (cf. Fig.~\ref{fig1}a). Up to at least $L\leq3\,l_{\rm C, e-ph}$, which corresponds to $L\approx 20\mu$m in our experiment, we observe a sizable dependence of $r_{\rm H}$ on the length of the constriction, see the solid line in Fig.~\ref{fig4}d. This dependence is described by the eqs.~(\ref{solution_r}) and (\ref{solution_alpha}) and contains two parameters associated with the e-ph relaxation length in the leads ($r_{\rm L, e-ph}\propto l_{\rm L, e-ph}$) and in the constriction ($l_{\rm C, e-ph}$). Also shown by the lower dashed line is the resistance of the constriction, $r_{\rm C}\propto L$, which crosses the dependence $r_{\rm H} (L)$ at about $L\approx 1.5\,l_{\rm C, e-ph}$ in this device. Finally, the upper dashed line illustrates the approximation (\ref{solution_short}), which adequately captures the physics at $L<0.5\,l_{\rm C, e-ph}$. \change{Note, that the analytic solution (\ref{solution_r}-\ref{solution_short}) is only applicable in a situation when the cross-section  $A_{\rm L}$ of the leads is constant over a few e-ph lengths $l_{\rm L, e-ph}$. In case the shape of the leads is such that $A_{\rm L}$ varies the trends of Fig.~\ref{fig4}d will still hold qualitatively.}

We conclude this section by formulating a realistic strategy for a calibration of the thermal bias in a general experiment with non-ideal heaters. We envision an experiment which measures a quantity proportional to the applied thermal bias $\delta T$, such that the evolution of the signal with the length of the heater constriction can be used for the purpose of calibration. The strategy contains the following three steps: 
 
\begin{itemize}
	\item Design a device with at least two metallic heaters in the form of a narrow constriction of a cross-section $A_{\rm C}$ connected to the leads of a much larger cross-section $A_{\rm L}\gg A_{\rm C}$. The length $L_i$ of the $i$-th constriction should vary substantially among the heaters. Ideally, $L_i$ should span the range of a few e-ph relaxation lengths, which according to our experiment corresponds roughly to a $\sim 10\,\mu$m scale for \remove{devices} \change{Au/Ni metallic bilayer} on a $\rm SiO_2$ substrate at low temperatures. 
	
	\item Measure the thermal response of the device with respect to all heaters in the linear response regime. For instance, in TE measurements this corresponds to $V_{\rm TE}^i=S\delta T^i=S(r_{\rm H}^{i}I_{\rm H})^2/8\mathcal{L}T_{\rm bath}$ for the $i$-th heater, where $S$ is the Seebeck coefficient of the device. 
	
	\item Fit the experimental functional dependence of $r_{\rm H}^{i}$ on $L_i$ using the eqs.~(\ref{solution_r}) and (\ref{solution_alpha}).
		Use the obtained fit parameters $r_{\rm L, e-ph}$ and $l_{\rm C, e-ph}$ for the absolute calibration of the thermal bias $\delta T^i$ and, thus, of the $S$ or another thermal response in question. We envision that the absolute accuracy of the measurement can be improved to within 10\%  with such a calibration, instead of about 100\% without it.
	\end{itemize}

}

\section*{Summary}

In summary, we achieved accurate thermal biasing of a nanoscale electronic device at low temperatures by means of a contact heating approach.  Using the average noise thermometry and InAs NW-based local noise sensing we quantified the non-equilibrium electronic energy distribution and the temperature in the center of a metallic diffusive constriction in dependence of bias current. Numerical and analytic calculations allowed us to quantify the heat balance and the role of the non-ideal leads of the heater constriction in the experiment. We presented a simple strategy how to design the metallic heaters capable of generating a predictable thermal bias at nanoscale. 

\section*{Acknowledgement}
 We acknowledge valuable discussion with A.I.~Kardakova and technical assistance of D.~Ercolani. Financial support from the SUPERTOP project, QUANTERA ERA-NET Cofound in Quantum Technologies \change{and from  the FET-OPEN project AndQC} are acknowledged. Measurements of the local EED in section~\textbf{\nameref{section3}} and numerical calculations were supported by the RFBR project 19-32-80037. Measurements of the local and average noise thermometry were supported by the RSF project 19-12-00326. Analytic calculations in section~\textbf{\nameref{section5}} were performed under the state task of the ISSP RAS.

\section*{Materials and Methods} 

NW devices were fabricated starting from gold catalyzed n-doped InAs NWs with typical length of 4\,$\mu$m and a diameter 85\,nm grown by chemical beam epitaxy~\cite{0268-1242-30-11-115012}. The carrier density of the InAs NWs derived by field effect measurements is about \remove{1}$\change{2}\times10^{18}\rm cm^{-3}$. Typical ohmic contact resistance in our devices is below $100\,\rm\Omega$, whereas the NW resistance is about $10\,\rm k\Omega$ per micrometer.  The metallic nanostructures were realized by electron beam lithography (EBL) process involving two stages. First, 280 nm thick PMMA 950 K resist was spin-coated and followed by a soft-bake at $170^{\circ}$\,C for 90 sec. The sample was then exposed for e-beam (10\,kV) writing. Ni/Au (10/120 nm) was deposited via thermal evaporation on the e-beam written pattern for lift-off. Prior to the Ni/Au deposition, the NWs were passivated using an ammonium polysulfide $\rm (NH_{4})_{2}S_{x}$ solution, which ensured the formation of low-resistance ohmic contacts. Second, an additional standard EBL process was performed to achieve a precise overlay (with an accuracy $\sim15\,\rm nm$) and intentionally double the Ni/Au thickness, in the lead areas (see Fig. 1d).

We performed most of the measurements in two $\rm ^{3}He$ inserts, with the samples immersed in liquid (at $T_{\rm bath}$= 0.5\,K) or in gas (at $T_{\rm bath}= 4.2\,\mathrm{K}$) and placed vertically face down. The EED data of Fig.~\ref{fig2}b (body) were obtained in a cryo-free dilution refrigerator, with the sample in vacuum and inside the metallic case. Here, the lowest achievable electronic temperature in equilibrium did not exceed 80\,mK, verified via noise thermometry. The shot noise spectral density was measured using home-made low-temperature amplifiers (LTamp) with a voltage gain of about 10 dB, input current noise of $\sim10^{-27}\, \mathrm{A^2/Hz}$ and dissipated power of $\sim 200\,\mathrm{\mu W}$. We used a resonant tank circuit at the input of the LTamp, see the sketch in Fig.~\ref{fig1}a, with a ground bypass capacitance of a coaxial cable and contact pads $\sim40\,\mathrm{pF}$, a hand-wound inductance of $\sim 5\,\mathrm{\mu H}$ and a load resistance of $10\,\mathrm{k\Omega}$. The output of the LTamp was fed into the low noise 75 dB total voltage gain room temperature amplification stage followed by a hand-made analogue filter and a power detector. The setup has a bandwidth of $\Delta f\sim 0.5\,$MHz around a center frequency of $\approx 11\,$MHz. A calibration was achieved by means of equilibrium Johnson-Nyquist noise thermometry. For this purpose we used a commercial pHEMT transistor connected in parallel with the device, that was depleted otherwise.


\bibliographystyle{iopart-num-lek}
\bibliography{lite}

\end{document}